\documentclass[10pt, conference]{IEEEtran}
\usepackage{cite}
\usepackage{amsmath,amssymb,amsfonts}
\usepackage{algorithmic}
\usepackage{graphicx}
\usepackage{textcomp}
\usepackage{xcolor}
\usepackage{booktabs}
\usepackage{diagbox}
\usepackage{import}
\usepackage{boxedminipage}
\usepackage{paracol}
\usepackage{xparse}
\usepackage[free-standing-units=true]{siunitx}
\def\BibTeX{{\rm B\kern-.05em{\sc i\kern-.025em b}\kern-.08em
    T\kern-.1667em\lower.7ex\hbox{E}\kern-.125emX}}

\newcommand{\real}[1]{\ensuremath{\text{Re}\left\{#1\right\}}}

\usepackage[acronym, toc, nonumberlist, ,nohypertypes={acronym}]{glossaries}
\newacronym{DNN}{DNN}{deep neural network}
\newacronym{CDR}{CDR}{coherent-to-diffuse power ratio}
\newacronym{DoA}{DoA}{direction-of-arrival}
\newacronym{TDoA}{TDoA}{time difference of arrival}
\newacronym{MSE}{MSE}{mean-squared error}
\newacronym{MLP}{MLP}{multilayer perceptron}
\newacronym{CRNN}{CRNN}{convolutional recurrent neural network}
\newacronym[plural=GPs,firstplural=Gaussian processes (GPs)]{GP}{GP}{Gaussian process}
\newacronym{GRU}{GRU}{gated recurrent unit}
\newacronym{RIR}{RIR}{room impulse response}
\newacronym{AWGN}{AWGN}{additive white Gaussian noise}
\newacronym{SNR}{SNR}{signal-to-noise ratio}
\newacronym{STFT}{STFT}{short-time Fourier transform}
\newacronym[plural=PSDs,firstplural=power spectral densities~(PSDs)]{PSD}{PSD}{power sprectral density}
\newacronym{AE}{AE}{absolute error}
\newacronym{MAE}{MAE}{mean-absolute error}
\newacronym{PE}{PE}{position error}
\newacronym{MPE}{MPE}{mean position error}
\newacronym{WASN}{WASN}{wireless acoustic sensor network}
\newacronym{OoR}{OoR}{out-of-range}
\newacronym{ASR}{ASR}{automatic speech recognition}
\newacronym{TDNN}{TDNN}{time delay neural network}
\newacronym{CNN}{CNN}{convolutional neural network}
\newacronym{WLS}{WLS}{weighted least squares}
\newacronym{LS}{LS}{least squares}
\newacronym{RANSAC}{RANSAC}{random sample consensus}

\makeglossaries

\NewDocumentCommand\myframedtext{ s O{.9\linewidth} m }{%
	\IfBooleanTF{#1}{\begin{figure*}}{\begin{figure}}%
			\centering%
			\fbox{\parbox{#2}{%
					#3%
			}}%
			\IfBooleanTF{#1}{\end{figure*}}{\end{figure}}}

\begin{document}

\title{Deep Neural Network based Distance Estimation for Geometry Calibration in Acoustic Sensor Networks}
\author{Tobias Gburrek${}^{1}$, Joerg Schmalenstroeer${}^{1}$, Andreas Brendel${}^{2}$, Walter Kellermann${}^{2}$, Reinhold Haeb-Umbach${}^{1}$ \\
\textit{${}^{1}$Department of Communications Engineering, Paderborn University, Germany} \\
\textit{${}^{2}$Multimedia Communications and Signal Processing, FAU Erlangen-Nuernberg, Germany} \\
\{gburrek, schmalen, haeb\}@nt.uni-paderborn.de, \{Andreas.Brendel, Walter.Kellermann\}@FAU.de
}
\maketitle

\begin{abstract}
We present an approach to deep neural network based~(DNN-based)\glsunset{DNN} distance estimation in reverberant rooms for supporting 
geometry calibration tasks in wireless acoustic sensor networks. 
Signal diffuseness information from acoustic signals is aggregated via the coherent-to-diffuse power ratio to obtain a distance-related feature, which is mapped to a source-to-microphone distance estimate by means of a \gls{DNN}. This information is then combined with direction-of-arrival estimates from compact microphone arrays to infer the geometry of the sensor network. Unlike many other approaches to geometry calibration, the proposed scheme does only require that the sampling clocks of the sensor nodes are roughly synchronized. In simulations we show that the proposed \gls{DNN}-based distance estimator generalizes to unseen acoustic environments and that precise estimates of the sensor node positions are obtained.

%To this end we utilize \gls{DoA} estimates from compact microphone arrays by which there is no need for synchronization between the nodes of the \gls{WASN}.
%But \gls{DoA} estimates do not maintain information about the scaling of the geometry.
%We tackle this issue by using \glspl{DNN} to get estimates for the distance between a signal source and a node of the \gls{WASN}.
%Therefore, we employ the \gls{CDR} as information source, which delivers distance related diffuseness information from acoustic signals recorded by the single nodes of the \gls{WASN}.
%Thus, our approach remains applicable to non-synchronous \glspl{WASN}.
%In simulations we show that our distance estimator generalizes well to numerous environments and delivers accurate estimates.
%Furthermore, the combined usage of \gls{DoA} estimates and distance estimates for geometry calibration provides precise results.

%To this end we combine \gls{DoA} estimates from compact microphone arrays with distance estimates originating from an acoustic signal diffuseness information, by which our approach becomes applicable to non synchronous sensor networks.
%Thereby, we focus on the application of \glspl{DNN} for estimation of the distance between a signal source and a node of the \gls{WASN} utilizing the \gls{CDR}.

\end{abstract}

\begin{IEEEkeywords}
DNN, CDR, acoustic distance estimation, geometry calibration
\end{IEEEkeywords}

% !TeX spellcheck = en_US
% !TeX encoding = UTF-8
% !TeX root = Gburrek2020.tex
% !TeX program = pdflatex

\vspace{-0.01cm}
\section{Introduction}

A \gls{WASN} consists of small devices called ''nodes'', which are connected via wireless links. Each node is equipped with memory, a wireless network interface, a processing unit and one or multiple microphones. \Glspl{WASN} are used in surveillance, human-machine interfaces and environmental monitoring tasks \cite{Bertrand2011}.
Distributing microphones in an environment comes with the promise that there is always a sensor close to each relevant sound source. Thus, \glspl{WASN} offer the potential of improved signal enhancement and acoustic localization capabilities, compared to a single compact microphone array. 

Acoustic source localization can, e.g., be used to steer a camera towards a moving speaker in a smart home scenario~\cite{Schmalen2009}. In such multi-modal setups the usage of a common coordinate system eases the process of data fusion.
Hence, knowledge of the position and orientation of the sensor nodes within a chosen coordinate system is required in these scenarios to provide absolute positioning information.
The process of determining the nodes' position and orientation is called geometry calibration. 

However, manual geometry calibration is a tedious task, in particular if the network consists of a large number of nodes, and any change in the setup asks for recalibration. Therefore, automatic geometry calibration from the observed acoustic signals is desirable, and, indeed, has been studied extensively, see \cite{PJHF16} for an overview.

%Nowadays, \glspl{WASN} consisting of spatially distributed nodes equipped with microphone arrays have several application areas, like smart homes or smart meeting rooms.
%In these scenarios they can fulfill many tasks, e.g., acoustic event detection or speech enhancement.
%\comment{Better example for task that needs absolute geometry}
%For some tasks, like source localization, the position and the orientation of the nodes has to be known.
%The process of finding this information is called geometry calibration. 
%Often, \glspl{WASN} contain a large number of nodes, so that a manual geometry calibration becomes a tedious task, and automatic procedures are preferred.

It appears natural to consider geometry calibration and sampling clock synchronization jointly, because correlation-based measures, such as the \gls{TDoA} of signals at different nodes, typically used to infer geometric relations among the nodes, require a synchronous network. However, such methods (e.g.,\cite{Ma2019,Jia2018})
often require additional information, such as the position of anchor sources \cite{Ma2019}, which may not be available in practice.

In this paper we take a different approach: rather than relying on a synchronous network or jointly estimating the geometry and the sampling clock offsets, we develop a technique which only needs a rough synchronization across sensor nodes.
We do, however, assume that each node has a synchronous microphone array of known topology instead of only a single microphone. 
The rough synchronization between the nodes is needed to match the observations made by the different nodes.

%In our scenario nodes are randomly distributed in a room. We can assume that each node has a synchronously sampling microphone array with a known shape (see \cite{Afifi2018marvelo} for a hardware example). However, the nodes are not synchronized and neither the positions of the nodes nor the positions of the acoustic sources are known.

Our proposed approach to geometry calibration is an extension of the method we presented in~\cite{DOA_calib_Jacob}, which utilizes \gls{DoA} estimates computed from the microphone array signals of each node.
But, \gls{DoA} information alone can only infer a ``relative'' geometry, lacking any absolute distances. Thus, the inferred geometry can be determined only up to an unknown scaling factor. In this contribution we propose to compute this scaling factor from source-to-sensor distance estimates gleaned from the acoustic properties of the microphone signals. 

In \cite{brendel_distributed_2019} and \cite{brendel_probabilistic_2019} it was shown that the \gls{CDR} can be utilized to estimate the distance between a microphone pair and an acoustic source. There, \glspl{GP} were used for \gls{CDR}-based distance estimation.
However, the \glspl{GP} were learned for a certain acoustic environment and generalization to unseen environments is expected to yield poor results.

We overcome this restriction by using \glspl{DNN}, which are trained on various acoustic environments such that both, the room characteristics and the distance, can be extracted from the \gls{CDR}. To further support the generalization of the learned model to varying acoustic environments, we employ the recently proposed R-vectors as additional input to the network, which are meant to capture the room properties~\cite{r_vec}.

The remainder of the paper is organized as follows:
In Sec.~\ref{sec:dist_est} the idea of \gls{CDR}-based distance estimation is reviewed, followed by the new \gls{DNN} approach in Sec.~\ref{subsec:DNN_dist_est}. Subsequently, geometry calibration using \glspl{DoA} and distance estimates is explained in Sec.~\ref{sec:Calib}. Finally, simulation results are presented in 
Sec.~\ref{sec:Sim} and some conclusions are drawn in Sec.~\ref{sec:Conclusions}.

\section{CDR-Based Distance Estimation}
\label{sec:dist_est}
We consider a microphone pair, which records a single acoustic source in a reverberant environment. The recorded signal can be decomposed into a coherent component and a diffuse part. 
The \gls{CDR} measures the power ratio of these components, which is related to the distance between the source and the microphones as shown in~\cite{brendel_distributed_2019}.
In \cite[Eq. 12]{CDR} a \gls{DoA}-independent estimator, yielding the estimate $\widehat{\mathrm{CDR}}(l, k)$, was derived, where $l$ indexes the time frame and $k$ the frequency bin, respectively. From $\widehat{\mathrm{CDR}}(l, k)$, the so-called diffuseness
\begin{align}
\mathrm{\hat{D}}(l, k) = \frac{1}{1 + \widehat{\mathrm{CDR}}(l, k)}
\end{align}
can be computed, which we will use in the following, because, unlike the \gls{CDR}, it is limited to the interval [0, 1].

%To achieve robustness against temporal inactivity of the acoustic source, it was proposed in~\cite{brendel_distributed_2019} to \changed{utilize the averaged diffuseness $\zeta$:
%	\begin{align}
%	\zeta = \frac{1}{N_t(k_{max} - k_{min})}\sum_{l=1}^{N_t}\sum_{k=k_{min}}^{k_{max}}\mathrm{\hat{D}}(l, k),
%	\end{align}
%	where $N_t$ corresponds to the number of used time frames and $[k_{min}, k_{max}]$ to the considered frequency interval.}
%%average the diffuseness $\mathrm{\hat{D}}(l, k)$ across a sufficiently large number of time frames and frequency bins, resulting in the averaged diffuseness $\zeta$.

To achieve robustness against temporal inactivity of the acoustic source, it was proposed in~\cite{brendel_distributed_2019} to average the diffuseness $\mathrm{\hat{D}}(l, k)$ across a sufficiently large number of time frames and frequency bins, resulting in the averaged diffuseness $\zeta$.

\subsection{GP-Based Distance Estimation}
\label{subsec:Drawbacks}
In prior works on \gls{CDR}-based distance estimation, e.g.,~\cite{brendel_distributed_2019} and \cite{brendel_probabilistic_2019}, it was proposed to use \gls{GP} regression trained on the averaged diffuseness $\zeta$ for distance estimation. 

\begin{figure}[htb]
	\centering
	\includegraphics[width=1.0\linewidth]{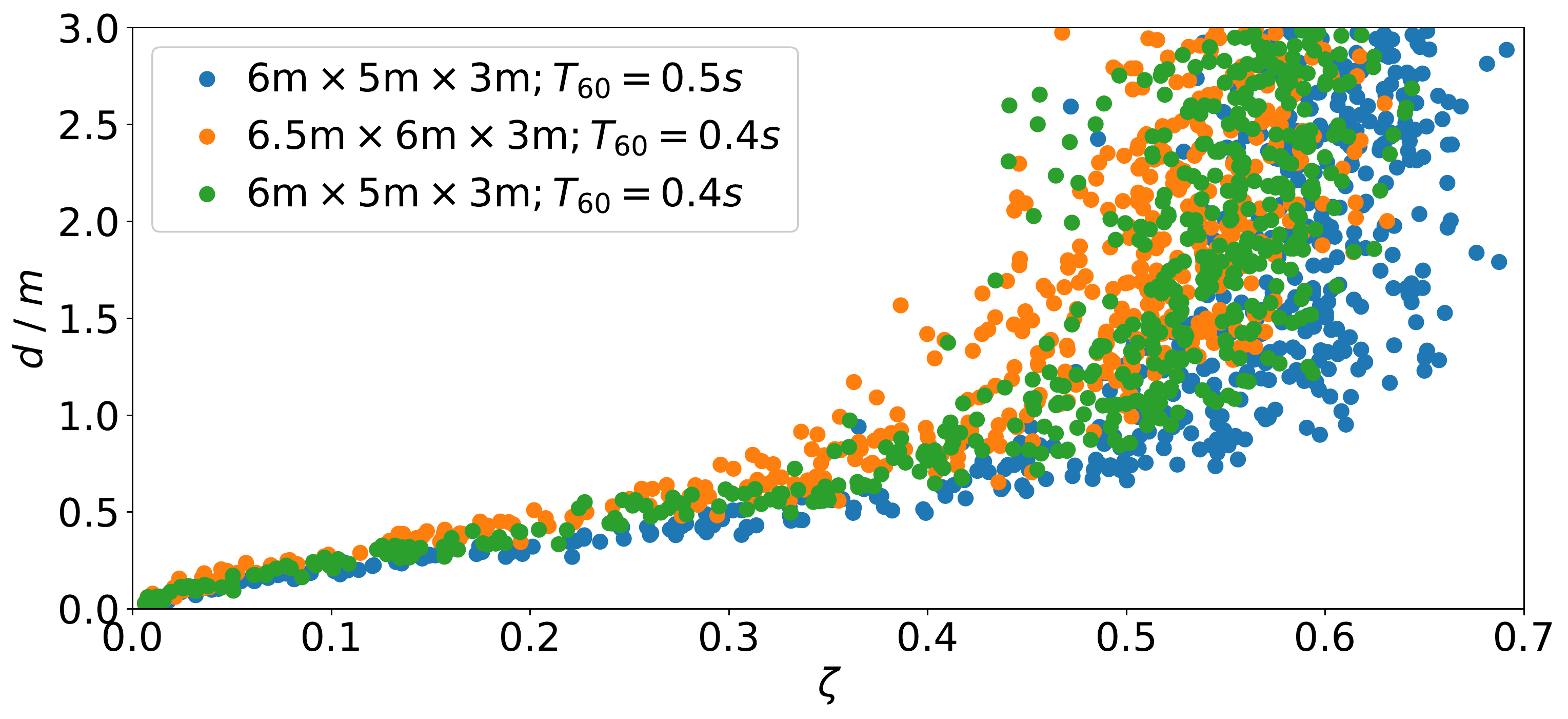}
	\caption{Dependency of the relationship between $\zeta$ and the distance $d$ on the acoustic environment: Each datapoint corresponds to a randomly drawn source-microphone constellation. The legend in the plot shows the dimensions of the considered rooms and the corresponding reverberation time $T_{60}$.}
	\label{fig:diff_dist_rel}
\end{figure}

Fig.~\ref{fig:diff_dist_rel} shows the relationship between the averaged diffuseness $\zeta$ and the distance $d$ for three exemplary rooms with different acoustic characteristics.
Obviously, this relationship strongly depends on the acoustic properties of the room, since
the energy decay of the coherent signal component is directly affected by the reverberation time $T_{60}$ and many other factors \cite{brendel_distributed_2019}. Consequently, a \gls{GP}, learned for a certain acoustic environment, will not generalize to other acoustic environments. 
Additionally, $\zeta$ tends for large values towards an asymptote, prohibiting the exact estimation of the regression function and, thus, degrading the performance also for smaller distances.

\section{DNN-Based Distance Estimation}
\label{subsec:DNN_dist_est}
\glspl{DNN} have many learnable parameters, which gives them an increased modeling power compared to \glspl{GP}. Hence, they may be able to take advantage of the additional information present in the high-resolution estimate $\mathrm{\hat{D}}(l, k)$ compared to the scalar value $\zeta$. Keeping all the information contained in the time-frequency pattern, the \gls{DNN} has the freedom to decide by itself, how to best combine this information, rather than defining this combination beforehand. Actually, we used a window consisting of several frames of the diffuseness as input feature.
Additionally, we enable the \gls{DNN} to learn room characteristics by presenting data from various acoustic environments. The underlying hypothesis is that this will allow the \gls{DNN} to map $\mathrm{\hat{D}}(l, k)$ to a distance estimate irrespective of the present room characteristics.

Fig.~\ref{fig:diff_dist_rel} shows that the variance of $\zeta$ grows as a function of the distance, which also holds for $\mathrm{\hat{D}}(l, k)$.
Therefore, to avoid unreliable measurements, we concentrate on small distances and exclude distances that are larger than an upper bound $r_{\text{max}}$.
This restriction is not detrimental for scaling geometries, since only one reliable distance estimate is sufficient and there is mostly a node of a \gls{WASN} close to each relevant sound source.

To handle the growing variance of $\mathrm{\hat{D}}(l, k)$ for larger distances we formulated distance estimation as a classification problem rather than a regression problem, by which small and large deviations from the ground truth distance (class) are penalized equally.
Thus, a distinction between the large distances, which are more tricky to be estimated correctly, is enforced.
When distance estimation is formulated as a regression problem this can be circumvented by the \gls{DNN} because the loss can be minimized by estimating an average distance for the larger distances.
The error due to the categorization of the distance into classes has a negligible effect on geometry scaling because this error is rather small (a few \si{\centi \metre}) compared to the inter-node distances (a few \si{\metre}). For distances larger than $r_{\text{max}}$ an additional class called \gls{OoR} is introduced.

We investigate two types of \glspl{DNN}, a simple \gls{MLP} and a \gls{CRNN}.
The architectures of the \glspl{DNN} are given in Tab.~\ref{tab:dist_est_mlp} and Tab.~\ref{tab:dist_est_crnn}, respectively.
Hereby, $B$ denotes the size of the mini-batches, $F$ the number of frequency bins, $T$ the number of time frames, $C$ the number of classes, and $R$ the dimension of the R-vector, which will be introduced later.

The major difference between the two types of \glspl{DNN} lies in the usage of temporal information.
When using the \gls{MLP}, we average the diffuseness over all time frames in the considered observation interval to obtain the input feature vector of the \gls{DNN}. 
Thus, all time information is discarded.
In contrast, the \gls{CRNN} is able to utilize temporal information contained in $\mathrm{\hat{D}}(l, k)$, e.g., information about the activity of the coherent source.
This will also be reflected by the simulation results in Sec.~\ref{sec:Sim}.

\begin{table}[htb]
	\caption{architecture of the mlp used for distance estimation: dropout with a probability of $0.5$ is used in all hidden layers.}
	\label{tab:dist_est_mlp}
	\centering
	\begin{tabular}{c | c}
		Block & Output shape \\
		\hline
		Diffuseness & $ B \times F$ \\
		optional: Concat R-vector & $ B \times (F + R)$ \\
		\hline
		$ 3 \times \text{fc}_{\text{ReLU}}(1024)$ & $B \times 1024$\\
		\hline
		$\text{fc}_{\text{Softmax}}(C)$ & $B \times C$
	\end{tabular}
\end{table}

\begin{table}[htb]
	\caption{architecture of the crnn used for distance estimation: each conv\{1,2\}d layer includes relu as activation and batch normalization. only the last output vector of the gru is forwarded to the classification network.}
	\label{tab:dist_est_crnn}
	\centering
	\begin{tabular}{c | c}
		Block & Output shape \\
		\hline
		Diffuseness & $ B \times 1 \times F \times T$ \\
		\hline
		$2 \times \text{Conv2d} (7\times3; 16)$ & $B \times 16 \times F \times T$\\
		MaxPool2d($4\times2$) & $B \times 16 \times \lfloor F/4 \rfloor \times \lfloor T/2 \rfloor$\\
		\hline
		$2 \times \text{Conv2d} (7\times3; 32)$ & $B \times 32 \times \lfloor F/4 \rfloor \times T$\\
		MaxPool2d($4\times2$) & $B \times 32 \times \lfloor F/16 \rfloor \times \lfloor T/4 \rfloor$\\
		Reshape & $B \times 32 \cdot \lfloor F/16 \rfloor \times \lfloor T/4 \rfloor$\\
		\hline
		$\text{Conv1d} (3; 512)$ & $B \times 512 \times \lfloor T/4 \rfloor$\\
		$\text{Conv1d} (3; 256)$ & $B \times 256 \times \lfloor T/4 \rfloor$\\
		\hline
		$2 \times \text{GRU} (256)$ & $B \times 256$\\
		\hline
		optional: Concat R-vector & $B \times (256 + R)$\\
		$\text{fc}_{\text{ReLU}}(256)$ & $B \times 256$\\
		$\text{fc}_{\text{Softmax}}(C)$ & $B \times C$
	\end{tabular}
\end{table}

Distance estimation can be further improved by utilizing R-vectors as additional input feature. R-vectors have been introduced in~\cite{r_vec} to capture information about the acoustic environment in automatic speech recognition. 

This idea can be transferred to distance estimation, whereby the R-vector is used to capture information about the current environment, e.g., the reverberation time $T_{60}$.
As shown in Tab.~\ref{tab:dist_est_mlp} and Tab.~\ref{tab:dist_est_crnn} the R-vector will be concatenated either with the input feature vector of the \gls{MLP} or the output of the \gls{GRU} layer of the \gls{CRNN}.

R-vectors correspond to the output of an intermediate layer of a \gls{DNN} trained to classify \glspl{RIR} from reverberated signal recordings.
%We refer to \cite{r_vec} for further information about R-vectors.
In~\cite{r_vec} it was suggested to use a \gls{TDNN} for R-vector extraction.
However, we decided to replace the \gls{TDNN} by a convolutional neural network for simplicity, whereby this decision was inspired by the x-vector extractor presented in~\cite{r_vec_nn}.
The architecture of our R-vector extractor can be found in Tab.~\ref{tab:r_vec}. 
We use the output of the first fully connected layer as R-vector.

\begin{table}[htb]
	\caption{architecture of the r-vector extractor: each conv1d layer includes relu as activation. batch normalization is used.}
	\label{tab:r_vec}
	\centering
	\begin{tabular}{c | c}
		Block & Output shape \\
		\hline
		MFCC & $ B \times 23 \times T$\\
		\hline
		$\text{Conv1d} (3; 128)$ & $B \times 128 \times T$\\
		$\text{Conv1d} (5; 128)$ & $B \times 128 \times T$\\
		$\text{Conv1d} (1; 128)$ & $B \times 128 \times T$\\
		\hline
		StatisticPool & $B \times 256$\\	
		\hline
		$\text{R-Vector} = \text{fc}_{\text{ReLU}}(512)$ & $B \times 512$\\
		$\text{fc}_{\text{ReLU}}(512)$ & $B \times 512$\\
		$\text{fc}_{\text{Softmax}}(C)$ & $B \times C$
	\end{tabular}
\end{table}

\section{Geometry Calibration}
\label{sec:Calib}
\begin{figure}[htb]
	\centering
	\def\svgwidth{1.0\columnwidth}
	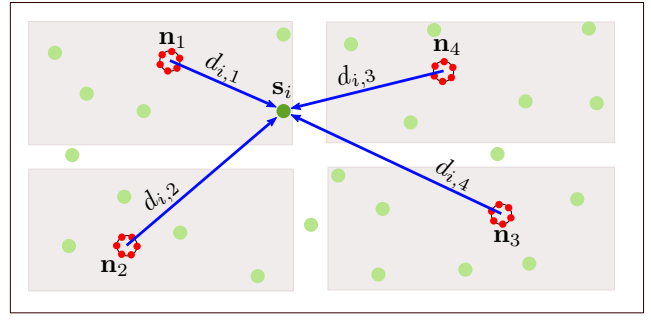
	\caption{Example of a random setup with four microphone arrays at positions $\mathbf{n}_j$, and highlighted $i$-th source position $\mathbf{s}_i$ (red dots: microphones; green dots: source positions; gray area: possible positions to randomly place nodes (microphone arrays); all nodes and sources have a minimum distance of \SI{0.5}{\metre} to the closest wall; \SI{1}{\metre} spacing between the gray areas; the dimensions of the room are drawn from [\SI{6}{\metre}, $\SI{7}{\metre}] \times [\SI{5}{\metre}, \SI{6}{\metre}]$; room height=\SI{3}{\metre})}
	\label{fig:calib_Setup}
\end{figure}

We simulated \mbox{two-dimensional} scenarios with $K$ nodes and $N$ independent acoustic sources, with only one source being active at any given time (see Fig.~\ref{fig:calib_Setup} for an example).
%An exemplary setup for $K{=}4$ can be found in Fig.~\ref{fig:calib_Setup}.

In order to determine the nodes' positions and orientations, we use the geometry calibration method introduced in \cite{DOA_calib_Jacob}, which also provides the position of the acoustic sources.
There, an objective function is defined which assesses the compatibility of the $K{\cdot}N$ \gls{DoA} estimates with an assumed geometry. This nonlinear objective function is iteratively minimized, see \cite{DOA_calib_Jacob} for details.
Additionally, we embed the calibration method into a similar \gls{RANSAC} method as the one described in~\cite{scale_inv} to be more robust against outliers in the \gls{DoA} estimates.

Due to the fact that this method only utilizes \gls{DoA} estimates, the optimization problem suffers from scale invariance \cite{scale_inv}. 
%In \cite{scale_inv} it is mentioned, that an additional equation, which defines the distance between two unknowns, or \gls{TDoA} estimates can be used to solve this issue. But, either prior knowledge (distance) or a synchronous \gls{WASN} (\gls{TDoA}) is required.
To avoid the trivial solution (all unknowns equal to zero), the following equality constraint, which relates all inter-node distances, is added to the optimization problem:
\begin{align}
	\sum_{i=1}^{K}\sum_{j=i}^{K} ||\mathbf{n}_i - \mathbf{n}_j||_2 = 1,
\end{align}
with  $\mathbf{n}_i$ and $\mathbf{n}_j$, $i, j\in \{1, 2, 3, 4\}$, denoting the node positions.
%Now, the optimization problem provides an estimate for the geometry, which is scaled by an unknown scaling factor $v$.

We use the estimated source-node distances to determine the unknown scaling factor $v$ of the calibration results, which arises from the introduced constraint.
As mentioned earlier, a single source-node distance estimate would ideally be sufficient.
%But, there are two sources for errors, which can be compensated, if more estimates will be used:
%First, the results of geometry calibration will not be perfect and, second, the distance estimates can be erroneous.
But better results are obtained if all available distance estimates are utilized.

The unknown $v\in \mathbb{R}^+$ is determined by scaling the \mbox{source-node} distances of the unscaled geometry to the distance estimates.
This results in the following weighted least squares problem
\begin{align}
	\label{eq:scaling_prob}
	\hat{v} = \underset{v}{\text{argmin}} \sum_{i=1}^{N} \sum_{j=1}^{K} w_{ij} \left(v || \hat{\mathbf{s}}_i - \hat{\mathbf{n}}_j||_2 - \hat{d}_{ij} \right)^2,
\end{align}
where $\hat{\mathbf{s}}_i$ denotes the unscaled estimate of the position of the \mbox{$i$-th} source, $\hat{\mathbf{n}}_j$ the unscaled estimate of the position of the \mbox{$j$-th} node, and $\hat{d}_{ij}$ the estimate of the distance between source $i$ and node $j$.
The weights $w_{ij}$ are introduced to account for the distance dependence of the variance of the distance estimates, see Fig.~\ref{fig:diff_dist_rel}. They are chosen to be: $w_{ij} = 1/|| \hat{\mathbf{s}}_i - \hat{\mathbf{n}}_j||_2$ .

The optimization in Eq.~\eqref{eq:scaling_prob} leads to 
\begin{align}
	\hat{v} = \frac{\sum_{i=1}^{N} \sum_{j=1}^{K} w_{ij} \: \hat{d}_{ij} \: || \hat{\mathbf{s}}_i - \hat{\mathbf{n}}_j||_2}{\sum_{i=1}^{N} \sum_{j=1}^{K}w_{ij} \: || \hat{\mathbf{s}}_i - \hat{\mathbf{n}}_j||_2^2}.
\end{align}
The properties of the sensor nodes used in the simulations were inspired by the hardware described in~\cite{Afifi2018marvelo}, where each node is equipped with a circular array that consists of six microphones. 
The two opposite microphones which are \SI{5}{\centi \metre} apart form a pair used for distance estimation, giving three distance estimates per array, which are combined by checking the consistency of the three estimates.
Using the microphones exhibiting the largest distance in an array is a reliable choice in practice (see, e.g.,~\cite{brendel_distributed_2019}).
If at least two estimates coincide, we select the corresponding estimate for geometry scaling and exclude the source-node pair otherwise ($w_{ij} = 0$).
%Otherwise, this source-node pair is excluded from geometry scaling ($w_{ij} = 0$).
Besides, we do not utilize the corresponding source-node pair for geometry scaling if at least one node provides the \gls{OoR} class.

\section{Simulation Results}
\label{sec:Sim}
Simulated data is used for the evaluation of our approach as well as for training the \glspl{DNN}.
We utilize the image source method~\cite{image_source} to simulate \glspl{RIR}, using the implementation of \cite{habets_rir}.
The \glspl{RIR} are used to reverberate the source signals, which can be either white Gaussian noise or speech, whereby the used speech samples are taken from the TIMIT database~\cite{timit}.
Due to additional physical effects that are not considered by the simulated data, e.g., directional sources, an adaptation of \gls{DNN}-based distance estimation to real data is expected to be beneficial for real microphone recordings and will be considered in future work.
%Due to the large difference between real data and simulated data, e.g., directed sources, an adaption of \gls{DNN}-based distance estimation to real data is necessary, which will be considered in future works.

The distance estimators are trained and evaluated on data sets, consisting of single source-node pairs, which are uniformly drawn from the room layout at a height of \SI{1.5}{\metre}. Due to the fact that the accuracy of distance estimates degrades if the source or the node is located in the vicinity of walls (see \cite{brendel_acoustic_2019}), a minimum distance of \SI{0.5}{\metre} to the closest wall is ensured for all nodes and sources. 

We use separate data sets for distance estimator training and R-vector extractor training, which we consider to be more realistic than using the same data sets for the training of both. In both data sets rooms with a height of \SI{3}{\metre} are considered.
The room dimensions are uniformly drawn from the set $[\SI{6}{\metre}, \SI{7}{\metre}] \times [\SI{5}{\metre}, \SI{6}{\metre}]$ for the distance estimator and from the set $[\SI{5}{\metre}, \SI{8}{\metre}] \times [\SI{4}{\metre}, \SI{7}{\metre}]$ for the R-vector extractor.
Besides, the reverberation time $T_{60}$ is uniformly drawn at random from [\SI{0.2}{\second}, \SI{0.5}{\second}] and [\SI{0.1}{\second}, \SI{0.6}{\second}], respectively.
We placed the sources such that the distance to the nodes is uniformly drawn from [\SI{0.03}{\metre}, \SI{3}{\metre}].
In the R-vector data set all sources are uniformly distributed in the room.
Both training data sets contain $10000$ source-node pairs, whereby additional $1000$ \gls{OoR} examples are added for the simulations corresponding to Tab.~\ref{tab:dist_est_snr_oor} and Tab.~\ref{tab:calib}.

In order to evaluate our approach to geometry calibration, we consider scenarios with a setup as depicted in Fig.~\ref{fig:calib_Setup}.
All scenarios consist of $K{=}4$ nodes and $N{=}30$ successively active sources, whereby each source corresponds to a \SI{3}{\second} long speech signal, which is generated as described before.
 
\subsection{Model Configuration}
All \glspl{DNN} are trained using Adam~\cite{Adam} with a mini-batch size of $B{=}32$ and a learning rate of $3{\cdot}10^{-4}$, whereby the distance is linearly quantized into $C{=}31$ classes plus one additional class for \gls{OoR}.
The short-time Fourier transform, which is needed to estimate $\widehat{\mathrm{CDR}}(l, k)$, uses a Blackman window with a length of \SI{25}{\milli\second} and \SI{10}{\milli\second} shift.
Additionally, we estimate the power spectral densities, used for \gls{CDR} estimation, by recursive averaging with forgetting factor $\lambda{=}0.95$, as described in \cite{brendel_distributed_2019}.
Furthermore, $\mathrm{\hat{D}}(l, k)$ is calculated for all frequencies between \SI{125}{\hertz} and \SI{3.5}{\kilo\hertz}, which corresponds to the frequency range, where speech has significant power.

\subsection{Distance Estimation}
We first evaluate the proposed distance estimators, and use the \gls{MAE} as performance metric
\begin{align}
	e_{AE} = \frac{1}{M}\sum_{m=1}^{M}|\hat{d}_m - d_m|.
\end{align} 
Here, $d_m$ denotes the ground truth distance and $\hat{d}_m$ the distance estimate.
The evaluation set contains $10000$ source-node constellations, which results in $M{=}30000$ source-microphone-pair constellations.

\begin{figure}[tb]
    \centering
    \includegraphics[width=1.\linewidth]{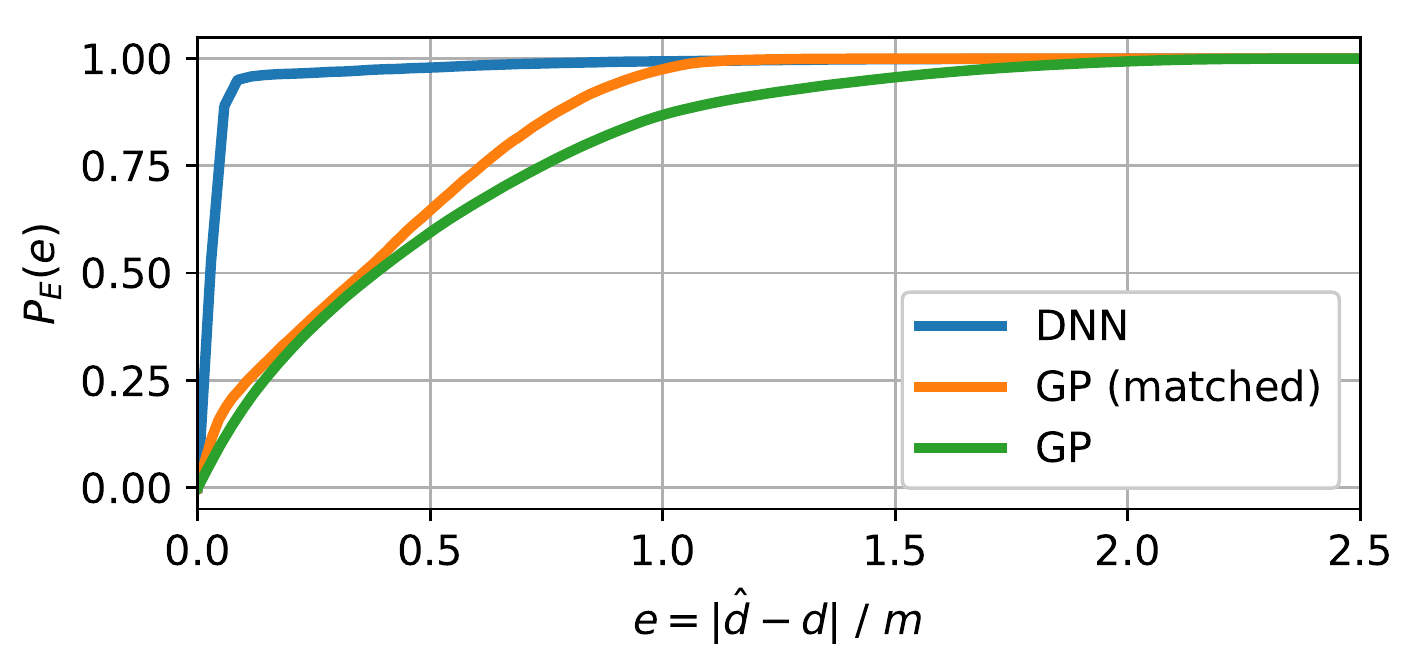}
    \caption{Cumulative distribution function of the error $P_E(e)$ of \gls{CRNN}-based and \gls{GP}-based distance estimates: The \gls{GP} (zero-mean prior and $\gamma$-exponential covariance function~\cite{brendel_probabilistic_2019}) is trained on a single acoustic environment and tested either on the same environment (matched) or on the evaluation data set that contains multiple environments. Speech is used as source signal.}
    \label{fig:comparison}
\end{figure}
Fig.~\ref{fig:comparison} shows a comparison of \gls{GP}-based and \gls{DNN}-based distance estimation. 
It can be seen that the proposed approach outperforms the \gls{GP}-based method, even so if the \gls{GP} is applied to the acoustic room characteristics, on which it was trained.
%This can be explained on the basis of Fig.~\ref{fig:diff_dist_rel}.
%Thus, it is not possible to learn a correct regression function for the considered acoustic characteristics.
%Besides, our approach is able to be applied on various acoustic environments.
%\glspl{GP} do not have this property.

\begin{table}[b]
	\caption{performance of the distance estimator for different types of source signals and different dnn architectures}
	\label{tab:dist_est_architecture}
	\centering
	\begin{tabular}{c c c c c}
		\toprule
		& &  &  \multicolumn{2}{c}{\textbf{MAE} / m} \\		
		\textbf{Architecture} & \textbf{Diffuseness} & \textbf{R-vector} &\textbf{Noise} &\textbf{Speech}\\
		\midrule
		\gls{MLP} &  & \checkmark & 0.176 & 0.151\\
		\gls{MLP} & \checkmark &  & 0.119 & 0.148\\
		\gls{MLP} & \checkmark & \checkmark & 0.064 & 0.070\\
		\gls{CRNN} & \checkmark & & 0.087 & 0.055 \\
		\gls{CRNN}  & \checkmark & \checkmark & 0.062 & 0.052\\
		\bottomrule
	\end{tabular}
\end{table}
Tab.~\ref{tab:dist_est_architecture} provides results for distance estimation using different input signals and \gls{DNN} architectures.
It becomes obvious that the best results can be achieved, when a \gls{CRNN} is used, which is able to utilize temporal information.
%This holds especially for speech as input signal.
Additionally, R-vectors, which contain distance information by itself (see first row of Tab.~\ref{tab:dist_est_architecture}), are helpful to reduce the error in all cases.
Nevertheless, R-vectors have a diminishing effect, when a \gls{CRNN} is used and speech is the input signal.
This means that the diffuseness contains already enough information about the environment.
Moreover, better results can be achieved when speech is used instead of white Gaussian noise.
We hypothesize that the correlation properties of the diffuseness resulting from speech support the learning process of the convolutional layers and, thus, are beneficial for gathering information about  distances and environments.

\begin{table}[thb]
	\setlength{\tabcolsep}{4pt}
	\caption{influence of snr and oor detection on distance estimation: 2500 oor examples are added to the evaluation set. the crnn~(diffuseness + r-vector) is applied to speech.}
	\label{tab:dist_est_snr_oor}
	\centering
	\begin{tabular}{c c c c c c}
		\toprule
		\textbf{SNR/\decibel} & \textbf{Fusion} &\textbf{\# Discards} &\textbf{MAE}/m &\textbf{F}$_\mathbf{1}$-\textbf{score (\gls{OoR} Detection)}\\
		\midrule
		30 & & - & 0.046 & 91.44\%\\
		30 &\checkmark  & 148& 0.033 & 94.90\%\\
		20 &\checkmark & 166 & 0.034 & 94.67\%\\
		10 &\checkmark & 308 & 0.044 & 91.86\%\\
		5 &\checkmark & 745 & 0.062 & 86.00\%\\
		\bottomrule
	\end{tabular}
\end{table}

The influence of sensor noise, which is simulated by adding white Gaussian noise to the reverberated signal, and the introduced \gls{OoR} class can be seen in Tab.~\ref{tab:dist_est_snr_oor}.
To generate the corresponding training data, integer values in the range from \SI{5}{\decibel} to \SI{30}{\decibel} are randomly chosen for the \gls{SNR}.
Obviously, our approach is robust against a wide range of sensor noise levels.
Furthermore, the fusion of the three distance estimates per node improves the performance. % of the distance estimator.
%Furthermore, adding sensor noise during training can be seen as data augmentation, which has a positive effect on the accuracy of the distance estimates.

\subsection{Geometry Calibration}
Next, we examine the geometry calibration performance.
It is to be mentioned that all results, which are provided by our geometry calibration method, are given relative to the position and orientation of a reference node.
Thus, the calibration results are matched to the ground truth geometry by a rigid body transformation for evaluation.
For \gls{DoA} estimation, the complex Watson kernel method~\cite{DOA_est} is used.

The \gls{MPE} of the nodes' positions is used as metric
\begin{align}
	e_{PE} = \frac{1}{4 M} \sum_{m=1}^{M}\sum_{j=1}^{4}||\hat{\mathbf{n}}_{j,m}-\mathbf{n}_{j,m}||_2.
\end{align}

Tab.~\ref{tab:calib} shows the \gls{MPE} of the calibration results for different values of $T_{60}$ for $M{=}100$ scenarios.
Noticeably, the \gls{MPE} increases for larger $T_{60}$ values.
This is caused by the degradation of the \gls{DoA} estimates in more reverberant environments (see~\cite{DOA_est}).
Additionally, the distance estimation errors influenced the calibration results only marginally.

\begin{table}[thb]
	\caption{mpe of the scaled geometry calibration results}
	\label{tab:calib}
	\centering
	\begin{tabular}{| c | c | c | c | c |}
		\hline
		\diagbox[]{Distances}{$T_{60}/\text{ms}$} & 200 & 300 & 400 & 500\\
		\hline
		Ground truth & 0.080\metre & 0.134\metre & 0.187\metre & 0.213\metre\\
		\hline
		Estimates (CRNN) & 0.084\metre & 0.14\metre1 & 0.197\metre & 0.228\metre\\
		\hline
	\end{tabular}
\end{table}

\section{Conclusions}
\label{sec:Conclusions}
In this paper we proposed a \gls{DNN}-based distance estimator. It takes acoustic signal diffuseness information and, optionally, R-vectors to capture acoustic properties of the room, as input and predicts the distance between an acoustic source and a recording node. The distance estimates are combined with \gls{DoA} estimates to infer the positions and orientations of the sensor nodes of a \gls{WASN}.
%Moreover, an approach to geometry calibration, which is applicable to non-synchronous \glspl{WASN},
Simulations have shown that the distance estimator provides estimates,  which exhibit an average error that is smaller than \SI{6.5}{\centi \metre} under various acoustic conditions, and enables accurate geometry calibration results.

\section*{Acknowledgment}
This work was supported by DFG under contract no {\small $<$SCHM 3301/1-2$>$} and {\small $<$KE 890/10-2$>$} within the framework of the Research Unit FOR2457 ``Acoustic Sensor Networks''.

\bibliographystyle{IEEEtran}
\bibliography{p1,library,asn_publications}

\end{document}